\documentclass[runningheads]{llncs}
\usepackage[hyphens]{url}
\usepackage{hyperref}
\usepackage[hyphenbreaks]{breakurl}

\usepackage{ifthen}
\usepackage{url}
\usepackage{amssymb}
\usepackage{amsmath}
\usepackage{import}
\usepackage{xparse}
\usepackage{amsmath}
\usepackage[usenames,dvipsnames]{xcolor}
\usepackage{xspace}
\usepackage{datatool}
\usepackage{glossaries}

\providecommand\m[1]{\ensuremath{#1}\xspace}
\renewcommand{\m}[1]{\ensuremath{#1}\xspace}

	\newcommand{\lrule}{\leftarrow}
	\newcommand{\cause}{\stackrel{c}{\lrule}}

	\newcommand{\voc}{\m{\Sigma}}

	\newcommand{\struct}{\m{I}}

	\NewDocumentCommand\inter{g+g}{%
	  \IfNoValueTF{#1}
	    {\struct}
	    {\m{#1^{#2}}}}

	\renewcommand{\int}{\m{\mathbb{Z}}}

	\NewDocumentCommand\subs{g+g}{%
	  \IfNoValueTF{#1}
	    {\m{/}}
	    {\m{#1/ #2}}}

\newcommand{\ouracronym}[3]{%
	\newacronym{#1}{#2}{#3}
	\expandafter\newcommand\csname #1\endcsname{\gls{#1}\xspace}%
}
	\ouracronym{FO}{FO}{first-order logic}
	\ouracronym{PC}{PC}{propositional calculus}
	\ouracronym{MX}{MX}{Model Expansion}
	\ouracronym{MO}{MO}{Model Optimization}
	\ouracronym{ASP}{ASP}{Answer Set Programming}
	\ouracronym{TP}{TP}{Theorem Proving}
	\ouracronym{LP}{LP}{Logic Programming}
	\ouracronym{CP}{CP}{Constraint Programming}
	\ouracronym{FP}{FP}{Functional Programming}
	\ouracronym{KR}{KR}{Knowledge Representation}
	\ouracronym{CSP}{CSP}{Constraint Satisfaction Problem}
	\ouracronym{SMT}{SMT}{SAT Modulo Theories}
	\ouracronym{KBS}{KBS}{knowledge base system}
	\ouracronym{NNF}{NNF}{Negation Normal Form}
	\ouracronym{FNNF}{FNNF}{Flat Negation Normal Form}
	\ouracronym{DefNNF}{DefNNF}{Definition Negation Normal Form}
	\ouracronym{DEFNF}{DEFNF}{Definition Normal Form}
	\ouracronym{CDCL}{CDCL}{Conflict-Driven Clause-Learning}
	\ouracronym{WFS}{WFS}{Well-Founded Semantics}
	\ouracronym{LCG}{LCG}{Lazy Clause Generation}
	\ouracronym{AEL}{AEL}{Autoepistemic Logic}
	\ouracronym{OEL}{OEL}{Ordered Epistemic Logic}
	\ouracronym{AFT}{AFT}{Approximation Fixpoint Theory}

	\makeatletter
	\def\ifenv#1{
	\def\@tempa{#1}%
	\def\@ttempa{#1*}%
	\ifx\@tempa\@currenvir
	\expandafter\@firstoftwo
	\else
	\expandafter\@secondoftwo
	\fi
	}
	\makeatother

	\newcommand{\ddrule}[4]{\ensuremath{#1 \leftarrow #2 & \{#3\} & #4}}
	\newcommand{\drule}[2]{\ensuremath{#1 & \leftarrow & #2}}

	\newcommand{\darule}[4]{\ensuremath{#1 \leftarrow #2 & \{#3\} & #4}}
	\newcommand{\arule}[2]{\ensuremath{#1 \, &\leftarrow \, #2}}

	\newcommand{\LNDRule}[2]{
	\ifenv{array}
	{\drule{#1}{#2}}
	{ \ifenv{align}
		{\arule{#1}{#2}}
		{\ifenv{align*}
		{\arule{#1}{#2}}
		{ERROR: using LDRule in unsupported environment: \@currenvir}
		}
	}
	}

	\newcommand{\LDRule}[4]{
	\ifenv{array}
	{\ddrule{#1}{#2}{#3}{#4}}
	{ \ifenv{align}
		{\darule{#1}{#2}{#3}{#4}}
		{\ifenv{align*}
		{\darule{#1}{#2}{#3}{#4}}
		{ERROR: using LDRule in unsupported environment: \@currenvir}
		}
	}
	}

	\NewDocumentCommand\LRule{m+g+g+g}{%
		\IfNoValueTF{#2}%
		{#1.&}{%
		\IfNoValueTF{#3}
		{\LNDRule{#1}{#2.}}
		{\LDRule{#1}{#2.}{#3}{#4}}%
		}
	}

	\NewDocumentCommand\CLRule{m+g}{%
	\ifenv{array}
	{\cdrule{#1}{#2}}
	{ \ifenv{align}
		{\carule{#1}{#2}}
		{\ifenv{align*}
			{\carule{#1}{#2}}
			{ERROR: using CLRule in unsupported environment: \@currenvir}
		}
	}
	}

	\NewDocumentCommand\carule{m+g}{%
		\IfNoValueTF{#2}
			{\ensuremath{#1.}}
			{\ensuremath{#1 \, &\cause \, #2}}}
	\NewDocumentCommand\cdrule{m+g}{%
		\IfNoValueTF{#2}
			{\ensuremath{#1.}}
			{\ensuremath{#1 & \cause & #2}}}

	\newcommand{\algrule}[4]{
	\hbox{{#1}:}& 
	\quad #2 ~\longrightarrow~ #3 
	\hbox{~ if } #4\\
	}

	\newcommand{\AlgoRule}[4]{
	\ifenv{array}
	{\algrule{#1}{#2}{#3}{#4}}
		{ERROR: using AlgoRule in unsupported environment: \@currenvir}
	}

	\newcommand{\ignore}[1]{}

	\newboolean{nocomments}
	\setboolean{nocomments}{false}

	\newboolean{commentmargin}
	\setboolean{commentmargin}{true}

	\newcommand{\namedcomment}[3]{%
		\ifthenelse{\boolean{nocomments}}%
		{}
		{
			\ifthenelse{\boolean{commentmargin}}%
				{ {\color{#3} \marginpar{\color{#3}\sc #2}#1}  }
				{  {\color{#3} {\sc #2}: #1}  }
		}%
	}
	\newcommand{\mnamedcomment}[3]{\ifthenelse{\boolean{nocomments}}{}{{\marginpar{ \color{#3}{\sc #2}:#1}}}}

	\usepackage{soul}

\usepackage[normalem]{ulem} 
\makeatletter
\font\uwavefont=lasyb10 scaled 700
\def\spelling{\bgroup\markoverwith{\lower3.5\p@\hbox{\uwavefont\textcolor{Red}{\char58}}}\ULon}
\def\grammar{\bgroup\markoverwith{\lower3.5\p@\hbox{\uwavefont\textcolor{LimeGreen}{\char58}}}\ULon}
\def\phrasing{\bgroup\markoverwith{\lower3.5\p@\hbox{\uwavefont\textcolor{RoyalBlue}{\char58}}}\ULon}

\newcommand\remove{\bgroup\markoverwith{\textcolor{red}{\rule[0.5ex]{2pt}{0.4pt}}}\ULon}
\makeatother

\usepackage{booktabs}
\usepackage{cleveref} 

\usepackage{tikz}
\usetikzlibrary{backgrounds}
\makeatletter

\tikzset{%
  fancy quotes/.style={
    text width=\fq@width pt,
    align=justify,
    inner sep=1em,
    anchor=north west,
    minimum width=\linewidth,
  },
  fancy quotes width/.initial={.8\linewidth},
  fancy quotes marks/.style={
    scale=7,
    text=white,
    inner sep=0pt,
  },
  fancy quotes opening/.style={
    fancy quotes marks,
  },
  fancy quotes closing/.style={
    fancy quotes marks,
  },
  fancy quotes background/.style={
    show background rectangle,
    inner frame xsep=0pt,
    background rectangle/.style={
      fill=gray!15,
      rounded corners,
    },
  }
}

\newenvironment{fancyquotes}[1][]{%
\noindent
\tikzpicture[fancy quotes background]
\node[fancy quotes opening,anchor=north west] (fq@ul) at (0,0) {``};
\tikz@scan@one@point\pgfutil@firstofone(fq@ul.east)
\pgfmathsetmacro{\fq@width}{\linewidth - 2*\pgf@x}
\node[fancy quotes,#1] (fq@txt) at (fq@ul.north west) \bgroup}
{\egroup;
\node[overlay,fancy quotes closing,anchor=east] at (fq@txt.south east) {''};
\endtikzpicture}
  \newcommand{\Sch}{\mathcal{S}}
\usepackage{tikz}

\usepackage{verbatim}
\usetikzlibrary{arrows.meta,%
                petri,%
                topaths}%

\newcommand\dom{\m{\Delta}}
\usepackage{stmaryrd}
\newcommand{\iexpr}[2]{\llbracket #1 \rrbracket^{#2}}
\newcommand{\semi}[1]{\iexpr{#1}I}

\newcommand{\sem}[1]{\iexpr{#1}G}

\newcommand{\sems}[1]{\llbracket #1 \rrbracket^G_\star}

\usepackage{amssymb,amsmath}
\newcommand\shacl{SHACL\xspace}
\newcommand\owl{OWL\xspace}
\newcommand\closed{\m{\mathit{closed}}}

\usepackage{graphicx}
\setboolean{nocomments}{false}

\begin{document}
\mainmatter              
\title{SHACL: A Description Logic in Disguise%
}

\titlerunning{SHACL: A Description Logic in Disguise}  
%
\author{Bart Bogaerts\inst{1}%
\orcidID{0000-0003-3460-4251}
\and
Maxime Jakubowski\inst{2,1}%
\orcidID{0000-0002-7420-1337}
\and
Jan Van den Bussche\inst{2}%
\orcidID{0000-0003-0072-3252}
}
\authorrunning{B. Bogaerts et al.}
%
\tocauthor{Bart Bogaerts, Maxime Jakubowki, and Jan Van den Bussche}

\institute{Vrije Universiteit Brussel, Brussels, Belgium  \\
\email{\{bart.bogaerts,maxime.jakubowski\}@vub.be}\\
 \and
Universiteit Hasselt, Hasselt, Belgium
\email{\{maxime.jakubowski,jan.vandenbussche\}@uhasselt.be}}

\maketitle              

\begin{abstract}
  \shacl is a W3C-proposed language for expressing structural
  constraints on RDF graphs.  In recent years, SHACL's popularity has
  risen quickly.  This rise in popularity comes with questions related
  to its place in the semantic web, particularly about its relation to
  OWL (the de facto standard for expressing ontological information on
  the web) and description logics (which form the formal foundations
  of \owl).  We answer these questions by arguing that \emph{\shacl is
    in fact a description logic}.  On the one hand, our answer is
  surprisingly simple, some might even say obvious.  But, on the other
  hand, our answer is also controversial.  By resolving this issue
  once and for all, we establish the field of description logics as
  the solid formal foundations of SHACL.

\end{abstract}

\section{Introduction}

The Resource Description Framework (RDF \cite{rdf11primer}) is a
standard format for publishing data on the web.  RDF represents
information in the form of directed graphs, where labeled edges
indicate properties of nodes.  To facilitate more effective access and
exchange, it is important for a consumer of an RDF graph to know what
properties to expect, or, more generally, to be able to rely on
certain structural constraints that the graph is guaranteed to
satisfy.  We therefore need a declarative language in which such
constraints can be expressed formally.

Two prominent proposals in this vein have been ShEx \cite{shex} and
\shacl \cite{shacl}.  In both approaches, a formula expressing the
presence (or absence) of certain properties of a node (or its
neighbors) is referred to as a ``shape''.  In this paper, we adopt the
elegant formalization of shapes in SHACL proposed by Corman, Reutter
and Savkovic \cite{corman}.  That work has revealed a striking
similarity between \emph{shapes} and \emph{concept expressions},
familiar from description logics (DLs) \cite{dlhandbook}.

The similarity between \shacl and DLs runs even deeper when we account
for \emph{named shapes} and \emph{targeting}, which is the actual
mechanism to express constraints on an RDF graph using shapes.  A
\emph{shape schema} is essentially a finite list of shapes, where each
shape $\phi_s$ is given a name $s$ and additionally associated with a
target query $q_s$.  The shape--name combinations in a shape schema
specify, in DL terminology, an \emph{acyclyc TBox} consisting of all
the formulas
\[
s\equiv \phi_s
.\] 
Given an RDF graph $G$, this acyclic TBox determines a unique
interpretation of sets of nodes to shape names $s$.  We then say that
$G$ \emph{conforms} to the schema if for each query $q_s$, each node
$v$ returned by $q_s$ on $G$ satisfies $s$ in the extension of $G$.

Now interestingly, the types of target queries $q$ considered for this
purpose in \shacl as well as in ShEx, actually correspond to simple
cases of shapes $\phi_{q_s}$ and the actual integrity constraint thus
becomes

\[
\phi_{q_s}\sqsubseteq s
.\]

As such, in description logic terminology, a shape schema consists of
two parts: an acyclic TBox (defining the shapes in terms of the given
input graph) and a general TBox (containing the actual integrity
constraints).

\section{The Wedge} 
  
Despite the strong similarity between \shacl and DLs, and despite the
fact that in a couple of papers, \shacl has been formalized in a way
that is extremely similar to description logics
\cite{corman,andresel,leinberger}, this connection is not recognized
in the community.  In fact, some important stakeholders in \shacl
recently even wrote the following in a blog post explaining why they
use SHACL, rather than \owl:\\

\begin{fancyquotes}
 \owl was inspired by and designed to exploit 20+ years of research in Description Logics (DL).  This is a field of mathematics that made a lot of scientific progress right before creation of \owl. I have no intention of belittling accomplishments of researchers in this field. However, there is little connection between this research and the practical data modeling needs of the common real world software systems. \hfill --- \hfill \cite{BlogNoOwl}
\end{fancyquotes}\\

\noindent
thereby suggesting that SHACL and DLs are two completely separated worlds and as such contradicting the introductory paragraphs of this paper. 
On top of that, SHACL is presented by some stakeholders
\cite{knoflook} as an alternative to the Web ontology language
\owl \cite{owldl}, which is based on
the description logic SROIQ
\cite{sroiq}.

This naturally begs the question: which misunderstanding is it that
drives this wedge between communities? How can we explain this
discrepancy from a mathematical perspective (thereby patently ignoring
strategic, economic, social, and other aspects that play a role).

\section{SHACL, OWL, and Description Logics} 
Our answer is that there are two important differences between \owl
and SHACL that deserve attention.  These differences, however, do not
contradict the central thesis of this paper, which is that \emph{SHACL
  is a description logic}.

\begin{enumerate}
\item The first difference is that \textbf{in \shacl, the data graph
    (implicitly) represents a first-order interpretation, while in
    \owl, it represents a first-order theory (an ABox)}.  Of course,
  viewing the same syntactic structure (an RDF graph) as an
  interpretation is very different from viewing it as a theory.  While
  this is a discrepancy between OWL and SHACL, theories as well as
  interpretations exist in the world of description logic and as such,
  this view is perfectly compatible with our central thesis.  There
  is, however, one caveat with this claim that deserves some
  attention, and that is highlighted by the use of the world
  ``implicitly''.  Namely, to the best of our knowledge, it is never
  mentioned that the data graph simply represents a standard
  first-order interpretation, and it has not been made formal what
  \emph{exactly} the interpretation is that is associated to a graph.
  Instead, \shacl's language features are typically evaluated
  \emph{directly} on the data graph.  There are several reasons why we
  believe it is important to make this translation of a graph into an
  interpretation \emph{explicit}.
  \begin{itemize}
  \item This translation makes \emph{the assumptions SHACL makes about
      the data} explicit.  For instance, it is often informally stated
    that ``SHACL uses closed-world assumptions''
    \cite{ShaclOWLCompared}; we will make this statement more precise:
    SHACL uses closed-world assumptions with respect to the relations,
    but open-world assumptions on the domain.
  \item Once the graph is eliminated, we are in familiar territory. In
    the field of description logics a plethora of language features
    have been studied. It now becomes clear how to add them to \shacl,
    if desired. The 20+ years of research mentioned in
    \cite{BlogNoOwl} suddenly become directly applicable to \shacl.
  \end{itemize}
  
\item The second difference, which closely relates to the first, is
  that \textbf{\owl and \shacl have a different (default) inference
    task}: the standard inference task at hand in OWL is
  \emph{deduction}, while in \shacl, the main task is validation of
  RDF graphs against shape schemas.  In logical terminology, this is
  evaluating whether a given interpretation satisfies a theory (TBox),
  i.e., this is the task of \emph{model checking}.

  Of course, the fact that a different inference task is typically
  associated with these languages does not mean that their logical
  foundations are substantially different.  Furthermore, recently,
  other researchers
  \cite{leinberger,shaclsatsouth,DBLP:journals/corr/abs-2108-13063}
  have started to investigate tasks such as \emph{satisfiability} and
  \emph{containment} (which are among the tasks typically studied in
  DLs) for SHACL, making it all the more obvious that the field of
  description logics has something to offer for studying properties of
  SHACL.
\end{enumerate}

In the next section, we develop our formalization of \shacl, building
on the work mentioned above.  Our formalization differs form existing
formalizations of \shacl in a couple of small but important ways.
First, as we mentioned, we explicitly make use of a first-order
interpretation, rather than a graph, thereby indeed showing that
\shacl is in fact a description logic.  Second, the semantics for
\shacl we develop would be called a ``natural'' semantics in database
theory \cite{ahv_book}: variables always range over the universe of
all possible nodes.  The use of the natural semantics avoids an
anomaly that crops up in the definitions of Andre\c sel et al.\
\cite{andresel}, where an ``active-domain'' semantics is adopted
instead, in which variables range only over the set of nodes actually
occurring in the input graph.  Unfortunately, such a semantics does
not work well with constants. The problem is that a constant mentioned
in a shape may or may not actually occur in the input graph.  As a
result, the semantics adopted by Andre\c sel et al.\ violates familiar
logic laws like De Morgan's law.  This is troublesome, since automated
tools (and humans!\@) that generate and manipulate logic formulas may
reasonably and unwittingly assume these laws to hold.  Also other
research papers (see Remark~\ref{remark:leinberger}) contain flaws
related to not taking into account nodes that \emph{do not} occur in
the graph.  This highlights the importance of taking a logical
perspective on \shacl.

A minor caveat with the natural semantics is that decidability of
validation is no longer totally obvious, since the universe of nodes
is infinite.  A solution to this problem is well-known from relational
databases \cite[Theorem 5.6.1]{ahv_book}. Using an application of
solving the first-order theory of equality, one can reduce, over
finite graphs, an infinite domain to a finite domain, by adding
symbolic constants \cite{4rus,hs_domind}.  It turns out that in our
case, just a single extra constant suffices.
  
In this paper, we will not give a complete syntactic translation of
\shacl shapes to logical expressions.  In fact, such a translation has
already been developed by Corman et al.~\cite{corman}, and was later
extended to account for all \shacl features by
Jakubowski~\cite{Maxime}.  Instead, we show very precisely how the
data graph at hand can be viewed as an interpretation, and that after
this small but crucial step, we are on familiar grounds and know well
how to evaluate expressions.

As already mentioned before, our formalization of SHACL differs in a
couple of ways from existing work.  These design choices are grounded
in true SHACL: with each of them we will provide actual SHACL
specifications that prove that SHACL validators indeed behave in the
way we expected.  All our examples have been tested on three SHACL
implementations: Apache Jena
SHACL\footnote{\url{https://jena.apache.org/documentation/shacl/index.html}}
(using their Java library) TopBraid
SHACL\footnote{\url{https://shacl.org/playground/}} (using their Java
library as well as their online playground), and
Zazuko\footnote{\url{https://shacl-playground.zazuko.com/}} (using
their online playground). The raw files encoding our examples (SHACL
specifications and the corresponding graphs) are available
online.\footnote{\url{https://vub-my.sharepoint.com/:f:/g/personal/bart_bogaerts_vub_be/Eicv10DwSnVEnT0BWNwEW8QBFuQjYTbwYYct1WYrkoefKQ?e=XhE8o0}.}

All our SHACL examples will assume the following prefixes are defined:
\begin{verbatim}
@prefix ex: <http://www.example.org/> .
@prefix sh: <http://www.w3.org/ns/shacl#> .
\end{verbatim}

\newcommand\eq{\m{\mathit{eq}}}
\newcommand\disj{\m{\mathit{disj}}}
\newcommand{\geqn}[3]{\geq_{#1}#2.#3}
\newcommand\concept{\phi}

 \newcommand\defin{\m{D}}
 \newcommand\targets{\m{T}}

\section{SHACL: The Logical Perspective} \label{secdefs}

In this section of the paper we begin with the formal development.  We
define shapes, shape schemas, and validation.  Our point of departure
is the treatment by Andre\c sel et al.\ \cite{andresel}, which we
adapt and extend to our purposes.

From the outset we assume three disjoint, infinite universes $N$, $S$,
and $P$ of \emph{node names}, \emph{shape names}, and \emph{property
  names}, respectively.\footnote{In practice, node names, shape names,
  and property names are IRIs \cite{rdf11primer}, hence the
  disjointness assumption does not hold.  However, this assumption is
  only made for simplicity of notation.}

We define \emph{path expressions} $E$ and \emph{shapes} $\phi$ by the
following grammar:
\begin{align*}
  E & ::=  p \mid p^- \mid E \cup E \mid E \circ E \mid E^* \mid E?\\
  \concept &::= \top \mid s \mid \{c\} \mid \concept \land \concept \mid
  \concept \lor \concept \mid \neg \concept \mid
  {}\geqn{n}{E}{\concept} \mid  \eq(p,E)\mid \disj(p,E)\mid \closed(Q)
\end{align*}
where $p$, $s$, and $c$ stand for property names, shape names, and
node names, respectively, $n$ stands for nonzero natural numbers, and
$Q$ stands for finite sets of property names.  In description logic
terminology, a node name $c$ is a \emph{constant}, a shape name is a
\emph{concept name} and a property name is a \emph{role name}.

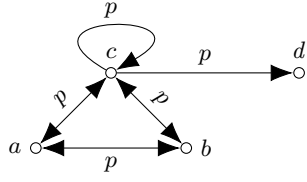
\begin{figure}[t]
\centering
\begin{tikzpicture}
\tikzstyle{grnode}=[draw,circle,fill=white,minimum size=4pt,
                            inner sep=0pt]
\draw (0,0) node [grnode] (a) [label=left:$a$] {};
 \draw (1,1) node [grnode] (c) [label=above:$c$] {};
  \draw (2,0) node [grnode] (b) [label=right:$b$] {};
  
 \draw (3.5,1) node [grnode] (d) [label=above:{$\mathit{d}$}] {};

\draw [{Latex[length=3mm]}-{Latex[length=3mm]}] (a) -- (c) node [sloped,midway,above ] {\small{$p$}};
\draw [{Latex[length=3mm]}-{Latex[length=3mm]}] (c) -- (b) node [sloped,midway,above ]{\small{$p$}};
\draw [{Latex[length=3mm]}-{Latex[length=3mm]}] (a) -- (b) node [sloped,midway,below] {\small{$p$}};
\draw [-{Latex[length=3mm]}] (c) edge[loop above,looseness=30,in=30,out=150]node{\small{$p$}} (c);

\draw [-{Latex[length=3mm]}] (c) -- (d) node [sloped,midway,above] {\small{$p$}};

\end{tikzpicture}
\caption{An example graph to illustrate language features of \shacl. }\label{fig:ex}
\end{figure}

As we will formalize below, every property/role name evaluates to a
binary relation, as does each path expression.  In the path
expressions, $p^-$ represents the inverse relation of $p$, $E\circ E$
represents composition of binary relations, $E^*$ the
reflexive-transitive closure of $E$ and $E?$ the reflexive closure of
$E$.  As we will see, shapes (which represent unary predicates) will
evaluate to a subset of the domain.  The three last expressions are
probably the least familiar. Equality ($\eq(p,E)$) means that there
are outgoing $p$-edges (edges labeled $p$) exactly to those nodes for
which there is a path satisfying the expression $E$ (defined
below). Disjointness ($\disj(p,E)$) means that there are \emph{no}
outgoing $p$-edges to which there is also a path satisfying $E$. For
instance in the graph in Figure~\ref{fig:ex}, $\eq(p,p^*)$ would
evaluate to $\{c\}$, since $c$ is the only node that has direct
outgoing $p$-edge to all nodes that are reachable using only
$p$-edges, and $\disj(p,p^-)$ would evaluate to $\{d\}$ since $d$ is
the only node that has no symmetric $p$-edges.  Closedness is also a
typical \shacl feature: $\closed(Q)$ represents that there are no
outgoing edges about any predicates other than those in $Q$.  In our
example figure $\closed(\{p\})$ would evaluate to $\{a,b,c,d\}$ and
$\closed(\{q\})$ to the empty set.

\begin{remark}
  Andre\c sel et al.\ \cite{andresel} also have the construct
  $\forall E.\phi$, which can be omitted (at least for theoretical
  purposes) as it is equivalent to $\neg\geqn{1}{E}{\neg \phi}$.  In
  our semantics, the same applies to $\phi_1 \land \phi_2$ and
  $\phi_1 \lor \phi_2$, of which we need only one as the other is then
  expressible via De Morgan's laws.  However, here we keep both for
  the sake of our later Remark~\ref{demorgan}.  In addition to the
  constructors of Andre\c sel et al.\ \cite{andresel}, we also have
  $E?$, \disj, and \closed, corresponding to \shacl features that were
  not included there.  \qed
\end{remark}

A \emph{vocabulary} $\Sigma$ is a subset of $N\cup S \cup P$.  A path
expression or shape is said to be \emph{over} $\Sigma$ if it only uses
symbols from $\Sigma$.

On the most general logical level, shapes are evaluated in
\emph{interpretations}.  We recall the familiar definition: An
interpretation $I$ over $\Sigma$ consists of
\begin{enumerate}
\item a set $\dom^I$, called the \emph{domain} of $I$;
\item for each constant $c \in \Sigma$, an element
  $\semi c \in \dom^I$;
\item for each shape name $s \in \Sigma$, a subset $\semi s$ of
  $\dom^I$; and
\item for each property name $p \in \Sigma$, a binary relation
  $\semi p$ on $\dom^I$.
\end{enumerate}
On any interpretation $I$ as above, every path expression $E$ over
$\Sigma$ evaluates to a binary relation $\semi E$ on $\dom^I$, and
every shape $\phi$ over $\Sigma$ evaluates to a subset of $\dom^I$, as
defined in Tables \ref{tab:sempath} and \ref{tab:sem}.

\begin{table}[t]
  \centering
  \begin{tabular}{@{}ll@{}}
    \toprule
    $E$ & $\iexpr{E}{I}$ \\
    \midrule
    $p^-$ & $\{(a,b) \mid (b,a) \in \iexpr{p}{I}\}$\\
    $E_1 \cup E_2$ & $\iexpr{E_1}{I} \cup \iexpr{E_2}{I}$\\
    $E_1 \circ E_2$ & $\{(a,b) \mid \exists c: (a,c) \in \iexpr{E_1}{I}  \land (c,b) \in \iexpr{E_2}{I} \}$ \\
    $E^*$ &the reflexive-transitive closure of
            $\iexpr{E}{I}$\\
    $E?$ &  $\iexpr{E}{I} \cup \{(a,a)\mid a\in\dom^I\}$\\
    \bottomrule
  \end{tabular}
  \caption{Semantics of a path expression $E$ in an interpretation $I$ over \voc.} \label{tab:sempath}
\end{table}
\begin{table}
  \centering
  \begin{tabular}{@{}ll@{}}
    \toprule
    $\concept$ & $\iexpr{\concept}{I}$\\
    \midrule
    $  \top$ & $ \dom^I$ \\
    $\{c\}$ & $\{c^I\}$\\
    $\concept_1 \land \concept_2$ & $\iexpr{\concept_1}{I} \cap \iexpr{\concept_2}{I}$\\
    $\concept_1 \lor \concept_2$ & $\iexpr{\concept_1}{I} \cup \iexpr{\concept_2}{I}$\\
    $\neg \concept_1$ & $\dom^I \setminus \iexpr{\concept_1}{I}$\\
    $\geqn{n}{E}{\concept_1} $ & $\{a \in \dom^I \mid \sharp (\semi{\phi_1} \cap \iexpr{E}{I}(a)) \geq n\} $\\
    $\eq(p,E) $ & $\{a \in \dom^I \mid
                  \semi{p}(a) = \semi{E}(a)\}$ \\
    $\disj(p,E) $ & $\{a \in \dom^I \mid
                    \semi{p}(a) \cap \semi{E}(a) = \emptyset\}$ \\
    $\closed(Q) $ & $\{a \mid \semi{p}(a)=\emptyset \text{  for every } p\in\voc\setminus Q\} $\\
    \bottomrule
  \end{tabular}
  \caption{Semantics of a shape $\concept$ in an interpretation $I$ over \voc. For a set
    $X$, we use $\sharp X$ to denote its cardinality. For a binary
    relation $R$ and an element $a$, we use $R(a)$ to denote the
    set $\{b \mid (a,b) \in R\}$.}\label{tab:sem}
\end{table}

As argued above, we define a \emph{shape schema} $\Sch$ over \voc as a
tuple $(\defin,\targets)$, where
\begin{itemize}
\item \defin is an \emph{acyclic TBox} \cite{dlhandbook}, i.e., a
  finite set of expressions of the form $s\equiv \concept_s$ with $s$
  a shape name in \voc and $\concept_s$ a shape over \voc and
  where
  \begin{enumerate}
  \item each $s$ occurs exactly once as the left-hand-side of such an
    expression and
  \item the transitive closure of the relation $\{ (s,t)\mid t$ occurs
    in $\concept_s\}$ is acyclic.
  \end{enumerate}
\item $\targets$ is a TBox, i.e., a finite set of statements of the
  form $\phi_1 \sqsubseteq \phi_2$, with $\phi_1$ and $\phi_2$ shapes.
\end{itemize}

\newcommand\extend[2]{\m{#1\!\uparrow\!#2}}
\renewcommand\extend[2]{\m{#1\diamond#2}}

If $\Sch=(\defin,\targets)$ is a shape schema over \voc and $I$ an
interpretation over $\voc\setminus S$, then there is a unique
interpretation $\extend{I}{\defin}$ that agrees with $I$ outside of
$S$ and that satisfies $\defin$, i.e., such that for every expression
$s\equiv\concept_s \in \defin$,
$\iexpr{s}{\extend{I}{\defin}} =
\iexpr{\concept_s}{\extend{I}{\defin}}$.  We say that $I$
\emph{conforms to} $\Sch$, denoted by $I \models \Sch$, if
$\iexpr{\phi_1}{\extend{I}{\defin}}$ is a subset of
$\iexpr{\phi_2}{\extend{I}{\defin}}$, for every statement
$\phi_1 \sqsubseteq \phi_2$ in $T$.  In other words, $I$ conforms to
$\Sch$ if there exists an interpretation that satisfies $\defin\cup T$
that coincides with $I$ on $N\cup P$.

\begin{remark}
  In real SHACL, a shape schema is called a ``shapes graph''.  There
  are some notable differences between shapes graphs and our shape
  schemas.

  First, we take abstraction of some features of real SHACL, such as
  checking data types like numbers and strings.

  Second, in real SHACL, the left-hand side of an inclusion statement
  in $\targets$ is called a ``target'' and is actually restricted to
  shapes of the following forms: a constant (``node target'');
  $\exists r.\{c\}$ (``class-based target'', where $r$ is `rdf:type');
  $\exists r.\top$ (``subjects-of target''); or $\exists r^-.\top$
  (``objects-of target'').  Our claims remain valid if this syntactic
  restriction imposed.

  Third, in real SHACL not every shape name needs to occur in the
  left-hand side of a defining rule. The default that is taken in real
  SHACL is that shapes without a definition are \emph{always
    satisified}. On the logical level, this means that for every shape
  $s$ name that has no explicit definition, a definition
  $s\equiv \top$ is implicitly assumed. The example that illustrates
  that our chosen default indeed corresponds to actual SHACL.  \qed
\end{remark}

\begin{example}
  The following SHACL shape \texttt{ex:MyShape} states that all nodes
  with an \texttt{ex:r}-edge must conform to the \texttt{ex:NoDef} and
  \texttt{ex:AlsoNoDef} shapes which we do not define.
\begin{verbatim}
ex:MyShape a sh:NodeShape ;
    sh:and ( ex:NoDef ex:AlsoNoDef ) .
ex:MyShape sh:targetSubjectsOf ex:r .
\end{verbatim}
  In our formal notation, this shapes graph corresponds to the shape
  schema
  \begin{align*}
    & \mathtt{ex{:}MyShape} \equiv \mathtt{ex{:}NoDef} \land \mathtt{ex{:}AlsoNoDef} \\
    & \exists \mathtt{ex{:}r}.\top \sqsubseteq \mathtt{ex{:}MyShape}
  \end{align*}
  where the first line is the definition of \texttt{ex:MyShape}, and the
  second line its target.

  When validating a graph containing only the triple \texttt{ex:a ex:r
    ex:b} (as we will show later, this corresponds to an interpretation
  in which the property name \texttt{ex:r} has the interpretation
  $\{(\texttt{ex:a},\texttt{ex:b})\}$ and the interpretation of all
  other property names is empty), and thus targeting the node
  \texttt{ex:a}, it validates without violation. This supports our
  observation that \textbf{shapes without an explicit definition are
    assumed to be satisfied by all nodes (i.e., are interpreted as
    $\top$)}.
  
  To further strengthen this claim, if instead we consider the SHACL
  shapes graph
\begin{verbatim}
ex:MyShape a sh:NodeShape ;
    sh:not ex:NoDef .
ex:MyShape sh:targetSubjectsOf ex:r .
\end{verbatim}
  i.e., the shape schema

  \begin{align*}
    & \mathtt{ex{:}MyShape} \equiv \lnot \mathtt{ex{:}NoDef} \\
    & \exists \mathtt{ex{:}r}.\top \sqsubseteq
      \mathtt{ex{:}MyShape}
  \end{align*}
  validation on the same graph yields the validation error that ``node
  \texttt{ex:a} does not satisfy \texttt{ex:MyShape} since it has shape
  \texttt{ex:NoDef}''.  \qed
\end{example}

\section{From Graphs to Interpretations}
Up to this point, we have discussed the logical semantics of \shacl,
i.e., how to evaluate a \shacl expression in a standard first-order
interpretation.  However, in practice, SHACL is not evaluated on
interpretations but on RDF graphs.  In this section, we show precisely
and unambiguously how to go from a graph to a logical interpretation
(in such a way that the actual SHACL semantics coincides with what we
described above).  A \emph{graph} is a finite set of \emph{facts},
where a fact is of the form $p(a,b)$, with $p$ a property name and $a$
and $b$ node names.  We refer to the node names appearing in a graph
$G$ simply as the \emph{nodes} of $G$; the set of nodes of $G$ is
denoted by $N_G$.  A pair $(a,b)$ with $p(a,b) \in G$ is referred to
as an \emph{edge}, or a \emph{$p$-edge}, in $G$.  The set of $p$-edges
in $G$ is denoted by $\sem p$ (this set might be empty).

We want to be able to evaluate \emph{any} shape on \emph{any} graph
(independently of the vocabulary the shape is over).  Thereto, we will
unambiguously associate, to any given graph $G$, an interpretation $I$
over $N \cup P$ as follows:
\begin{itemize}
\item $\dom^I$ equals $N$ (the universe of all node names).
\item $\semi c$ equals $c$ itself, for every node name $c$.
\item $\semi p$ equals $\sem p$, for every property name $p$.
\end{itemize}
If $I$ is the interpretation associated to $G$, we use $\sem E$ and
$\sem \phi$ to mean $\semi E$ and $\semi \phi$, respectively.

RDF also has a model-theoretic semantics \cite{rdfsemantics}.  These
semantics reflect the view of an RDF graph as a basic ontology or
logical theory, as opposed to the view of an RDF graph as an
interpretation.  Since the latter view is the one followed by SHACL,
it is thus remarkable that SHACL effectively ignores the
W3C-recommended semantics of RDF.

\begin{remark} \label{demorgan}
  Andre\c sel et al.\ \cite{andresel} define $\sem \phi$ a bit
  differently.  For a constant $c$, they define $\sem{ \{c\}} = \{c\}$
  like we do.  For all other constructs, however, they define
  $\sem \phi$ to be $\semi \phi$, but with the domain of $I$ taken to
  be $N_G$, rather than $N$.  In that approach, if $c \notin N_G$,
  $\sem {\neg \neg \{c\}}$ would be empty rather than $\{c\}$ as one
  would expect.  For another illustration, still assuming
  $c \notin N_G$, $\sem{\neg(\neg \phi \land \neg \{c\})}$ would be
  $\sem\phi$ rather than $\sem\phi \cup \{c\}$, so De Morgan's law
  would fail. The next examples shows that actual \shacl
  implementations indeed coincide with our semantics. \qed
\end{remark}

\begin{example}
  The following SHACL shape \texttt{ex:MyShape} states that it cannot
  be so that the node \texttt{ex:MyNode} is different from itself
  (i.e., that it must be equal to itself, but specified with a double
  negation).
\begin{verbatim}
ex:MyShape a sh:NodeShape ;
    sh:not [ sh:not [ sh:hasValue ex:MyNode ] ] .
ex:MyShape sh:targetNode ex:MyNode .
\end{verbatim}
  In our formal notation, this shapes graph corresponds to the shape
  schema
  \begin{align*}
    & \mathtt{ex{:}MyShape} \equiv \neg\neg\{\mathtt{ex{:}MyNode}\} \\
    & \{\mathtt{ex{:}MyNode}\} \sqsubseteq \mathtt{ex{:}MyShape}
  \end{align*}
  Clearly, this shape should validate every graph, also graphs in
  which the node \texttt{ex:MyNode} is not present and it indeed does
  so in all SHACL implementations we tested.  This supports our
  \textbf{choice of the natural semantics}, rather than the active
  domain semantics of \cite{andresel}.  Indeed, in that semantics,
  this shape will never validate any graph because the right-hand side
  of the inclusion will be evaluated to be the empty set.  \qed
\end{example}

\begin{example}
  Another example in the same vein as the previous, to show that the
  \textbf{natural semantics} correctly formalizes is the one where
  \cite{andresel}'s semantics does not respect the De Morgan's laws,
  as follows:
\begin{verbatim}
ex:MyShape a sh:NodeShape ;
    sh:not [
        sh:and (
            [ sh:not [
                  sh:path ex:r ;
                  sh:minCount 1 ] ]
            [ sh:not [ sh:hasValue ex:MyNode ] ] ) ] .
ex:MyShape sh:targetNode ex:MyNode .
\end{verbatim}
  This shapes graph corresponds to the shape schema
  \begin{align*}
    & \mathtt{ex{:}MyShape} \equiv \neg ( \neg \exists \mathtt{ex{:}r}.\top \land \neg \{\mathtt{ex{:}MyNode}\}) \\
    & \{\mathtt{ex{:}MyNode}\} \sqsubseteq \mathtt{ex{:}MyShape}
  \end{align*}
  In the formalism of Andre\c sel et al.\ \cite{andresel}, this schema
  does not validate on graphs that do not mention the node
  \texttt{ex:MyNode}, but in our formalism (and all SHACL
  implementations), it does validate.  \qed
\end{example}

\begin{remark} \label{remark:leinberger}
  The use of active domain semantics has also introduced some errors
  in previous work.  For instance \cite[Theorem 1]{leinberger} is
  factually incorrect.  The problem originates with the notion of
  \emph{faithful assignment} introduced by Corman et al.~\cite{corman}
  and adopted by Leinberger et al.  This notion is defined in an
  active-domain fashion, only considering nodes actually appearing in
  the graph. For a concrete counterexample to that theorem, consider a
  single shape named $s$ defined as $\exists r.\top$, with target
  $\{b\}$. In our terminology, this means that
  \[
    \begin{array}{l}
      \defin=\{s\equiv\exists r.\top\}\text{, and}\\
      T = \{\{b\}\sqsubseteq s\}.
    \end{array}
  \]
  On a graph $G$ in which $b$ does not appear, we can assign $\{s\}$
  to all nodes from $G$ with an outgoing $r$-edge (meaning that all
  these nodes satisfy $s$ and no other shape (names)), and assign the
  empty set to all other nodes (meaning that all other nodes do not
  satisfy any shape).  According to the definition, this is a faithful
  assignment.  However, the inclusion $\{b\} \sqsubseteq s$ is not
  satisfied in the interpretation they construct from this assignment,
  thus violating their Theorem~1.  \qed
\end{remark}

The bug in \cite{leinberger}, as well as the violation of De Morgan's
laws will only occur in corner cases where the shape schema mentions
nodes that not occur in the graph.  After personal communications,
Leinberger et al.\ \cite{leinberger} included an errata section where
they suggest to fix this by demanding that (in order to conform) the
target queries do not mention any nodes not in the graph.  While
technically, this indeed resolves the issue (under that condition,
Theorem 1 indeed holds), this solution in itself has weaknesses as
well.  Indeed, shape schemas are designed to validate graphs not known
at design-time, and it should be possible to check conformance of
\emph{any} graph with respect to \emph{any} shape schema.  As the
following example shows, it makes sense that a graph should conform to
a schema in case a certain node does \emph{not} occur in the graph (or
does not occur in a certain context), and that --- contrary to the
existing \shacl formalizations --- the natural semantics indeed
coincides with the behaviour of \shacl validators in such cases.

\begin{example}\label{ex:luis}
  Consider a schema with $\defin=\emptyset$ and $T$ consisting of a single inclusion 
  \[\{\mathit{MarcoMaratea}\}\sqsubseteq \lnot \exists (author\circ
    venue).\{\mathit{LPNMR22}\},\] which states that Marco Maratea
  (one of the LPNMR PC chairs) does not author any LPNMR paper.  If
  Marco Maratea does not occur in the list of of accepted papers, this
  list should clearly\footnote{Technically, the standard is slightly
    ambiguous with respect to nodes not occurring in the data graph.}
  conform to this schema.  This example can be translated into actual
  SHACL as follows:
\begin{verbatim}
ex:NotAnAuthor a sh:NodeShape ;
    sh:not [
        a sh:PropertyShape ;
        sh:path (ex:author ex:venue) ;
        sh:qualifiedValueShape [ sh:hasValue ex:LPNMR22 ] ;
        sh:qualifiedMinCount 1 ] .
ex:NotAnAuthor sh:targetNode ex:MarcoMaratea .
\end{verbatim}
  where we simply give the name \texttt{ex:NotAnAuthor} to the shape
  that holds for all nodes that do not author any LPNMR paper and
  subsequently enforce that Marco Maratea satisfy this shape.  We see
  that indeed, in accordance with our proposed semantics, graphs
  without a node \texttt{ex:MarcoMaratea} validate with respect to
  this SHACL specification.  The fix in the erratum of Leinberger et
  al.\ \cite{leinberger}, on the other hand, specifies that this does
  not validate. \qed
\end{example}

The definition of $I$ makes --- completely independent of the actual
language features of SHACL --- a couple of assumptions explicit.
First of all, SHACL uses unique names assumptions (UNA): each constant
is interpreted in $I$ as a different domain element.  Secondly, if
$p(a,b)$ does not occur in the graph, it is assumed to be
\emph{false}. However, if a node $c$ does not occur anywhere in the
graph, it is not assumed to not exist: the domain of $I$ is infinite!
Rephrasing this: \shacl makes the Closed World Assumption (CWA) on
predicates, but not on objects.

\paragraph{Effective evaluation} Since the interpretation defined from
a graph has the infinite domain $N$, it is not immediately clear that
shapes can be effectively evaluated over graphs.  As indicated above,
however, we can reduce to a finite interpretation.  Let
$\Sigma\subseteq N\cup P$ be a finite vocabulary, let $\phi$ be a
shape over $\Sigma$, and let $G$ be a graph.  From $G$ we define the
interpretation $I_\star$ over $\Sigma$ just like $I$ above, except
that the domain of $I_\star$ is not $N$ but rather
$$ N_G \cup (\Sigma \cap N) \cup \{\star\}, $$ where $\star$ is an
element not in $N$.  We use $\sems \phi$ to denote
$\iexpr \phi {I_\star}$ and find:

\begin{theorem}
  \label{thm:sterreke}
  For every $x \in N_G \cup (\Sigma \cap N)$, we have
  $x \in \sem \phi$ if and only if $x \in \sems \phi$.  For all other
  node names $x$, we have $x \in \sem \phi$ if and only if
  $\star \in \sems \phi$.

  Hence, $I$ conforms to $\Sch$ if and only if $I_\star$ does. 
\end{theorem}

Theorem \ref{thm:sterreke} shows that conformance can be performed by
finite model checking, but other tasks typically studied in DLs are
not decidable; this can be shown with a small modification of the
proof of undecidability of the description logic $\mathcal{ALRC}$, as
detailed by Schmidt-Schau\ss~\cite{KLONE}.
\begin{theorem}\label{thm:undec}
  Consistency of a shape schema (i.e., the question whether or not
  some $I$ conforms to $\mathcal{S}$) is undecidable.
\end{theorem}
Following description logic traditions, decidable fragments of SHACL
have been studied already; for instance Leinberger et al.\
\cite{leinberger} disallow equality, disjointness, and closedness in
shapes, as well as union and Kleene star in path expressions.

\section{Related Work and Conclusion} \label{secrel}
Formal investigations of SHACL have started only relatively recently.
We already mentioned the important and influential works by Corman et
al.~\cite{corman} and by Andre\c sel et al.~\cite{andresel}, which
formed the starting point for the present paper.  The focus of these
papers is mainly on the extending the semantics to \emph{recursive}
\shacl schemas, which are not present in the standard yet, and which
we also do not consider in the current paper.

The connection between SHACL and description logics has also been
observed by several other groups of researchers
\cite{leinberger,shaclsatsouth,DBLP:journals/corr/abs-2108-13063,ahmetaj-expl-shacl}.
There, the focus is on typical reasoning tasks from DLs applied to
shapes, and on reductions of these tasks to decidable description
logics or decidable fragments of first-order logic.  In its most
general form, this cannot work (see Theorem \ref{thm:undec}), but the
addressed works impose restrictions on the allowed shape expressions.

Next to shapes, other proposals for adding integrity constraints to
the semantic web have been proposed, for instance by integrating them
in OWL ontologies \cite{aaai/TaoSBM10,DBLP:journals/ws/MotikHS09}.
There, the entire ontology is viewed as an incomplete database.

None of the discussed works takes the explicit viewpoint that a data
graph represents a standard first-order interpretation or that SHACL
validation is model checking.  We took this viewpoint and in doing so
formalized precisely how SHACL relates to the field of description
logics.  There are (at least) three reasons why this formalization is
important.  First, it establishes a bridge between two communities,
thereby allowing to exploit the many years of research in DLs also for
studying SHACL.  Second, our formalization of SHACL clearly separates
two orthogonal concerns:
\begin{enumerate}
\item \emph{Which information does a data graph represent?} This is
  handled in the translation of a graph into its \emph{natural
    interpretation}.
\item \emph{What is the semantics of language constructs?} This is
  handled purely in the well-studied logical setting.
\end{enumerate}

Third, as we showed above, our formalization corresponds closer to
actual SHACL than existing formalizations, respects well-known laws
(such as De Morgan's) and avoids issues with nodes not occurring in
the graph requiring special treatment. As such, we believe that by
rooting SHACL in the logical setting, we have devised solid
foundations for future studies and extensions of the language.  We
already build on the logical foundations of the current paper in our
work on extending the semantics to \emph{recursive} shape schemas
\cite{iclp/BogaertsJ21}, as well as in an analysis of the primitivity
of the different language features of SHACL \cite{icdt/BogaertsJV22}.
 
\bibliographystyle{splncs04}
\bibliography{krrlib, database-shortened, extra-refs}

\begin{thebibliography}{10}
\providecommand{\url}[1]{\texttt{#1}}
\providecommand{\urlprefix}{URL }
\providecommand{\doi}[1]{https://doi.org/#1}

\bibitem{ahv_book}
Abiteboul, S., Hull, R., Vianu, V.: Foundations of Databases. Addison-Wesley
  (1995)

\bibitem{ahmetaj-expl-shacl}
Ahmetaj, S., David, R., Ortiz, M., Polleres, A., Shehu, B., Simkus, M.:
  Reasoning about explanations for non-validation in {SHACL}. In: Proceedings
  of KR. pp. 12--21. IJCAI Organization (2021)

\bibitem{andresel}
Andre\c{s}el, M., Corman, J., Ortiz, M., Reutter, J., Savkovic, O., Simkus, M.:
  Stable model semantics for recursive {SHACL}. In: Proceedings of {WWW}. pp.
  1570--1580 (2020)

\bibitem{4rus}
Aylamazyan, A., Gilula, M., Stolboushkin, A., Schwartz, G.: Reduction of the
  relational model with infinite domains to the case of finite domains. Doklady
  Akademii Nauk SSSR  \textbf{286}(2),  308--311 (1986), in Russian

\bibitem{dlhandbook}
Baader, F., Calvanese, D., McGuiness, D., Nardi, D., Patel-Schneider, P.
  (eds.): The Description Logic Handbook. Cambridge University Press (2003)

\bibitem{iclp/BogaertsJ21}
Bogaerts, B., Jakubowski, M.: Fixpoint semantics for recursive {SHACL}. In:
  Formisano, A., Liu, Y., et~al. (eds.) Proceedings ICLP. Electronic
  Proceedings in Theoretical Computer Science, vol.~345, pp. 41--47 (2021)

\bibitem{icdt/BogaertsJV22}
Bogaerts, B., Jakubowski, M., Van~den Bussche, J.: Expressiveness of {SHACL}
  features. In: Olteanu, D., Vortmeier, N. (eds.) Proceedings of ICDT.
  vol.~220, pp. 15:1--15:16. Schloss Dagstuhl--Leibniz-Zentrum f\"ur Informatik
  (2022)

\bibitem{shex}
Boneva, I., Gayo, J.L., Prud'hommeaux, E.: Semantics and validation of shape
  schemas for {RDF}. In: Proceedings of {ISWC}. pp. 104--120 (2017)

\bibitem{corman}
Corman, J., Reutter, J., Savkovic, O.: Semantics and validation of recursive
  {SHACL}. In: Proceedings of {ISWC}. pp. 318--336 (2018), extended version,
  technical report
  \href{https://www.inf.unibz.it/krdb/tech-reports/}{KRDB18-01}

\bibitem{sroiq}
Horrocks, I., Kutz, O., Sattler, U.: The even more irresistible {SROIQ}. In:
  Proceedings of {KR}. pp. 57--67 (2016)

\bibitem{hs_domind}
Hull, R., Su, J.: Domain independence and the relational calculus. Acta
  Informatica  \textbf{31},  513--524 (1994)

\bibitem{Maxime}
Jakubowski, M.: Formalization of {SHACL}.
  \url{https://www.mjakubowski.info/files/shacl.pdf}, accessed: 2021-06-16

\bibitem{ShaclOWLCompared}
Knublauch, H.: {SHACL} and {OWL} compared.
  \url{https://spinrdf.org/shacl-and-owl.html}, accessed: 2021-06-16

\bibitem{leinberger}
Leinberger, M., Seifer, P., Rienstra, T., L{\"{a}}mmel, R., Staab, S.: Deciding
  {SHACL} shape containment through description logics reasoning. In:
  Proceedings of {ISWC}. pp. 366--383 (2020)

\bibitem{DBLP:journals/ws/MotikHS09}
Motik, B., Horrocks, I., Sattler, U.: Bridging the gap between {OWL} and
  relational databases. J. Web Semant.  \textbf{7}(2),  74--89 (2009)

\bibitem{owldl}
{OWL}~2 {W}eb ontology language: {S}tructural specification and
  functional-style syntax. W3C Recommendation (Dec 2012)

\bibitem{shaclsatsouth}
Pareti, P., Konstantinidis, G., Mogavero, F., Norman, T.J.: {SHACL}
  satisfiability and containment. In: Proceedings of {ISWC}. pp. 474--493
  (2020)

\bibitem{DBLP:journals/corr/abs-2108-13063}
Pareti, P., Konstantinidis, G., Mogavero, F.: Satisfiability and containment of
  recursive {SHACL}. Journal of Web Semantics  (2022).
  \doi{https://doi.org/10.1016/j.websem.2022.100721},
  \url{https://arxiv.org/abs/2108.13063}

\bibitem{BlogNoOwl}
Polikoff, I.: Why {I} don't use {OWL} anymore -- {Top Quadrant} blog.
  \url{https://www.topquadrant.com/owl-blog/}, accessed: 2021-06-04

\bibitem{rdf11primer}
{RDF} 1.1 primer. W3C Working Group Note (Jun 2014)

\bibitem{rdfsemantics}
{RDF} 1.1 semantics. W3C Recommendation (Feb 2014)

\bibitem{KLONE}
Schmidt{-}Schau{\ss}, M.: Subsumption in {KL-ONE} is undecidable. In:
  Proceedings of {KR}. pp. 421--431 (1989)

\bibitem{shacl}
Shapes constraint language ({SHACL}). W3C Recommendation (Jul 2017)

\bibitem{aaai/TaoSBM10}
Tao, J., Sirin, E., Bao, J., McGuinness, D.L.: Integrity constraints in {OWL}.
  In: Proceedings of {AAAI} (2010)

\bibitem{knoflook}
TopQuadrant: An overview of {SHACL}: A new {W3C} standard for data validation
  and modeling. \url{https://www.topquadrant.com/an-overview-of-shacl/} (2017),
  webinar slides

\end{thebibliography}

\end{document}
